\documentclass[dvips,12pt]{emulateapj-rtx4}
\usepackage[english]{babel}
\usepackage{amsmath}
\usepackage{amssymb}
\usepackage{url}
\usepackage[dvips]{color}

\usepackage[dvipdfm,hypertexnames=false]{hyperref} 

\newcommand{\ergs}{\rm\,erg\,s^{-1}}

\newcommand{\msun}{\,M_{\odot}}

\newcommand{\mdot}{\dot{M}}
\newcommand{\dr}{\Delta r}
\newcommand{\drr}{\Delta r_2}
\newcommand{\SAX}{SAX J1808.4--3658}
\newcommand{\rc}{$r_{\rm co}$}
\newcommand{\rin}{$r_{\rm m}$}

\def\simless{{\th \rlap{\raise 0.5ex\hbox{$\scriptstyle  {<}$}}
    {\lower 0.3ex\hbox{$\scriptstyle  {\sim}$}} \th }}  
\def\simgreat{{\th \rlap{\raise 0.5ex\hbox{$\scriptstyle  {>}$}}
    {\lower 0.3ex\hbox{$\scriptstyle  {\sim}$}} \th }}  
\def\th{\thinspace}
\def\ts{{\raise 0.3ex\hbox{$\scriptstyle {\th \sim \th }$}}}

\begin{document}

\title{1 Hz Flaring in the Accreting Millisecond Pulsar NGC 6440 X-2: Disk Trapping and Accretion Cycles}
\shorttitle{1 Hz modulation and Magnetic Outbursts}
\shortauthors{Patruno et~al.}

\author{
  Alessandro Patruno\altaffilmark{1,2}, Caroline D'Angelo\altaffilmark{1}},
\altaffiltext{1}{Astronomical Institute ``Anton Pannekoek,''
  University of Amsterdam, Science Park 904, 1098 XH Amsterdam, The Netherlands}
\altaffiltext{2}{ASTRON, the Netherlands Institute for Radio Astronomy, Postbus 2, 7990 AA Dwingeloo, the Netherlands}

\begin{abstract}

  The dynamics of the plasma in the inner regions of an accretion disk
  around accreting millisecond X-ray pulsars is controlled by the
  magnetic field of the neutron star. The interaction between an
  accretion disk and a strong magnetic field is not well-understood,
  particularly at low accretion rates (the so-called ``propeller
  regime''). This is due in part to the lack of clear observational
  diagnostics to constrain the physics of the disk-field
  interaction. Here we associate the strong $\sim 1$ Hz modulation
  seen in the accreting millisecond X-ray pulsar NGC 6440 X-2 with an
  instability that arises when the inner edge of the accretion disk is
  close to the corotation radius (where the stellar rotation rate
  matches the Keplerian speed in the disk). A similar modulation has
  previously been observed in another accreting millisecond X-ray
  pulsar (\SAX) and we suggest that the two phenomena are related and
  that this may be a common phenomenon among other magnetized
  systems. Detailed comparisons with theoretical models suggest that
  when the instability is observed, the interaction region between the
  disk and the field is very narrow -- of the order of 1 km. Modeling
  further suggests that there is a transition region ($\sim 1-10$ km)
  around the corotation radius where the disk-field torque changes
  sign from spin up to spin down. This is the first time that a direct
  observational constraint has been placed on the width of the
  disk-magnetosphere interaction region, in the frame of the
  trapped-disk instability model.
\end{abstract}



\section{Introduction}

The serendipitous discovery of the accreting millisecond X-ray pulsar
(AMXP) NGC 6440 X-2 has presented new challenges to our understanding
of magnetospheric accretion. This source was discovered in 2009, and
is, on average, the faintest known AMXP \citep{alt10}.  The neutron
star spins at 206 Hz and orbits around its white dwarf companion in
about 1 hr\citep{alt10}.  Between 2009 and 2011 it underwent 10
outbursts, all of which were unusually short (2-3 days) with a very
short recurrence time ($\sim 1$ month; \citealt{hei10}) and peak luminosities
of about $10^{36}\rm\,erg\,s^{-1}$. Subsequent
time-series analysis revealed a strong modulation at 1 Hz during an
outburst \citep{pat10}. This modulation is similar to that observed in
\SAX\, in the decay tail of a number of outbursts
(\citealt{van00b,pat09c}).

Both sources show the modulation at moderately low luminosities,
around what is known as the ``propeller regime''. Accretion disks
surrounding strong stellar magnetic fields are disrupted close to the
star, where the field is strong enough to control the gas
dynamics. The location where the disk ends is usually termed the
magnetospheric radius (\rin). If the star is spinning rapidly or the
accretion rate is sufficiently low, the magnetospheric radius will lie
outside the corotation radius, the point at which the Keplerian disk
rotates at the same rate as the star 
\begin{equation}
r_{\rm co} \equiv \left(GM_*/\Omega^2_*\right)^{1/3}.
\end{equation} The rapidly-spinning magnetic
field then acts as a centrifugal barrier that inhibits accretion onto
the star. It is usually assumed that the gas in the disk will then be
expelled from the system, much like a propeller \citep{ill75}.

Recent new work, however, has suggested this picture is incomplete.
\cite{spr93} and \cite{rap04} noted that in order for a large amount
of mass to be expelled from the system the inner edge of the disk must
be (at minimum) substantially outside the corotation radius. If not,
the interaction between the magnetic field and disk will prevent gas
from accreting but not inject enough energy to expel it from the
disk. Furthermore, the coupling between the magnetic field and disk
will transfer angular momentum from the star to the disk, which will
both spin down the star and change the radial gas density distribution
in the inner disk region. This disk solution, known as a ``dead
disk'', was first described by \citealt{siu77}, but has been
little-examined since then.

When a dead disk forms in the inner disk regions, \cite{dan10,dan12}
demonstrated that continued accretion from large radii will often
drive an accretion instability. The instability sets in when the inner
edge of the disk moves outside \rc, halting accretion onto the
star. As gas continues to accumulate from the outer regions of the
disk, the inner edge of the disk gradually overcomes the centrifugal
barrier created by the magnetic field and eventually crosses \rc. Once
the disk inner edge has moved inside \rc, the accumulated reservoir of
gas is depleted and the inner edge again moves outside \rc, allowing a
new cycle to begin.

Building on earlier work by \cite{spr93} (which did not incorporate
the dead disk solution), \cite{dan10,dan12} studied this instability
extensively, demonstrating that it in fact exists in two
different accretion regimes, one at very low accretion rates, and one at
higher fluxes compatible to those observed in some AMXPs at the onset
of the propeller stage. 

The dead disk model also makes a radically different prediction for
disks in quiescence from the standard one. In the conventional model,
the inner edge of the disk is a straightforward function of accretion
rate $\dot{M}$, moving outwards as $\dot{M}$ decreases. In contrast,
the dead disk model predicts that even at very low accretion rates the
inner edge of the disk will not move far outside \rc, and a
significant amount of gas will remain in the inner disk region even
when $\dot{M}\simeq0$. \citet{dan11} demonstrated that in this case a
`trapped disk' can develop, in which the disk inner edge stays trapped
close to \rc~and acts as an efficient sink for angular momentum,
spinning down the accretor considerably. This disk solution will also
likely alter the duration of individual outbursts, which in the
ionization disk instability model ( see e.g.~\citealt{las01} for a
review), is affected by both the surface density in the inner disk
regions and the feedback of radiation from accretion onto the central
object back onto the disk.

In this paper we present an analysis of the unusual outburst behaviour
seen in NGC 6440 X-2 and the 1 Hz modulation and argue that both are
manifestations of an unstable dead disk, following the model of
\citet{dan10}. Using data from the \textit{Rossi X-ray Timing Explorer} ({\it
  RXTE}) we identify the 1 Hz modulations in six different outbursts
and we study its spectral and timing properties, demonstrating that
these are consistent with coming from accretion rate modulation onto
the neutron star. We propose that the 1 Hz modulations seen in NGC 6440
X-2 and SAX J1808.4--3658 have the same origin and are a new and
unique diagnostic to study the disk-magnetosphere interaction.

\section{X-Ray Observations}
\label{sec:observations}

\begin{deluxetable}{llll}
\tabletypesize{\footnotesize}
\tablecolumns{4}
\tablewidth{0pt}
\tablecaption{\textit{RXTE} and \textit{Swift} observations of NGC 6440 X-2}
\tablehead{
  \colhead{Outburst Start} &
  \colhead{1 Hz mod.} &
  \colhead{Instrument}&
  \colhead{(Obs-Id)}
}
\startdata
04-Jun-2009 & NA & \textit{Swift}/XRT & 31421001\\ 
30-Jul-2009 & NA & \textit{RXTE}/PCA  & Bulge-Scan\\
\smallskip
\textbf{30-Aug-2009} & Y & \textit{RXTE}/PCA  & 94044-04-02-00\\
01-Oct-2009 & NA & \textit{RXTE}/PCA  & Bulge-Scan\\
28-Oct-2009 & NA & \textit{RXTE}/PCA  & Bulge-Scan\\
\textbf{21-Mar-2010} & Y & \textit{RXTE}/PCA  & 94315-01-12-00;\\
\smallskip
& & &  94315-01-12-02\\
\textbf{12-Jun-2010} & Y & \textit{RXTE}/PCA  & 94315-01-14-00\\
\textbf{04-Oct-2010} & Y & \textit{RXTE}/PCA  & 94315-01-25-00 \\
\textbf{23-Jan-2011} & Y & \textit{RXTE}/PCA  & 96326-01-02-00\\
\textbf{06-Nov-2011} & Y & \textit{RXTE}/PCA  & 96326-01-35-00\\
\enddata
\label{tab:rxteobs}
\tablecomments{The outbursts discussed in this paper are highlighted in bold
text.}
\end{deluxetable}

We used all pointed {\it RXTE} observations recorded with the
Proportional Counter Array (PCA;~\citealt{jah06}) that cover a time
range from August 30, 2009 (MJD~55073) to November 18, 2011
(MJD~55883).  We used the Standard 2 mode data (16-s time resolution)
to calculate the X-ray lightcurve.  The background
counts are subtracted and dead-time corrections are made. The
energy-channel conversion is done by using the pca\_e2c\_e05v04 table
provided by the \textit{RXTE} team. In order to correct for the gain changes
and the differences in effective area between the different PCUs, the
flux is normalized to the Crab Nebula values (see
\citealt{kuu94}; \citealt{van03}) that are closest in time but in the
same \textit{RXTE} gain epoch \citep{jah06}.  

For the timing analysis we used all GoodXenon and Event data, with a
time resolution of $2^{-20}$ and $2^{-13}$ s, respectively. The
GoodXenon data were rebinned to the same resolution of the Event data.
No background subtraction or dead-time correction was applied to the
data before calculating the power spectra. The Poissonian noise level was
determined by taking the average power between 3000 and 4000 Hz, a
region in the power spectra dominated by counting statistics
noise. This mean value was then subtracted from the power spectra.  The
background was calculated for each observation (Obs-Id) with
the FTOOL {\tt pcabackest} over the entire energy range. 

We used 256s-long segments to calculate power spectra with no energy
selection (i.e., using all 256 energy channels). The
frequency boundaries for the power spectra are therefore $1/256$ Hz
and 4096 Hz.  The power spectra were normalized in the rms
normalization \citep{van95} which gives the power density in units of
$\rm\,(rms/mean)^2\,Hz^{-1}$. Consistent with \citet{pat09c}, we define
the fractional rms amplitude between $\nu_1$ and $\nu_2$ as:
\begin{equation}
\rm\,rms\it\, = \left[\int^{\nu_2}_{\nu_1}P(\nu)d\nu\right]^{\rm\,1/2}
\end{equation} 
and calculate the errors from the dispersion of the data points in the power
spectra. 

To confirm the presence of coherent X-ray pulsations at the spin
frequency ($\approx 206$ Hz) of the neutron star, we barycentered
the photons arrival times with the FTOOL {\tt faxbary} using the
most precise \textit{Chandra} astrometric position
available~\citep{hei10} and the JPL DE-405 solar system ephemeris.  We
then generated dynamical power spectra with 256 s long data segments
and overlapping intervals of 64 s. As is common practice in coherent
timing studies of AMXPs, we here retain only the energy channels
between 5 and 37 (absolute channels).

Each power spectrum is inspected by eye to look for the presence of
the 1 Hz modulation which was already observed and reported on June
12th 2010~\citep{pat10}.  The 1 Hz modulation was observed in the
power spectrum of \SAX\, as a broad feature that can be described by one
or multiple Gaussians~\citep{pat09c}.  Therefore we follow the same
procedure here and fit the 1 Hz modulation in the power spectrum with
one or more Gaussian functions:
\begin{equation}
P(\nu) = A\rm\,exp\left[(\nu-\nu_0)^2/{\it{C}}^2\right]
\end{equation} 
where A is the amplitude of the modulation, $\nu_0$ the centroid frequency
 and the constant $C$ is related
to the FWHM via the expression:
\begin{equation}
FWHM = 2C\cdot [\rm\,ln(2)]^{1/2}.
\end{equation}
The choice of a Gaussian distribution rather than a Lorentzian distribution
(which is more typical in QPOs fitting) is made because the power spectrum
shows a rapid decline at lower frequencies, leading to a systematic
overestimate when fitting with a Lorentzian function. 

 \begin{figure}[!t]
  \begin{center}
    \rotatebox{0}{\includegraphics[width=1.05\columnwidth]{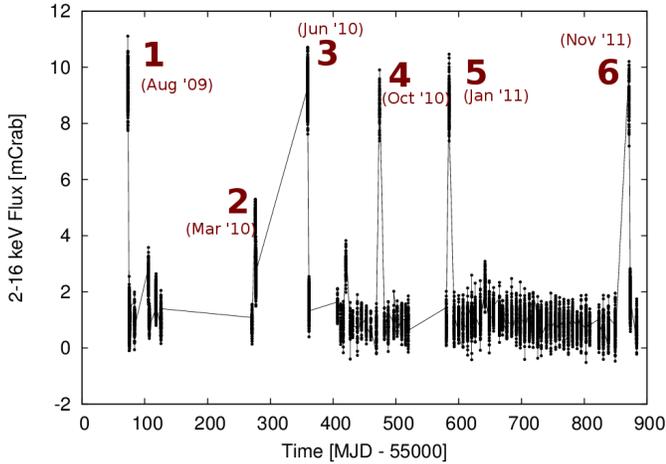}}
  \end{center}
  \caption{The 2-16 keV X-ray lightcurve of NGC 6440 X-2 from Aug 30, 2009
    (MJD~55073) until November 18, 2011. The plot shows the occurrence
    of six outbursts highlighted with a number. The 1 Hz modulation is
    detected in all six outbursts with the flux that is always above $\sim
    2$mCrab. \label{lc}}
\end{figure}

 \begin{figure}[!th]
  \begin{center}
    \rotatebox{-90}{\includegraphics[width=0.7\columnwidth]{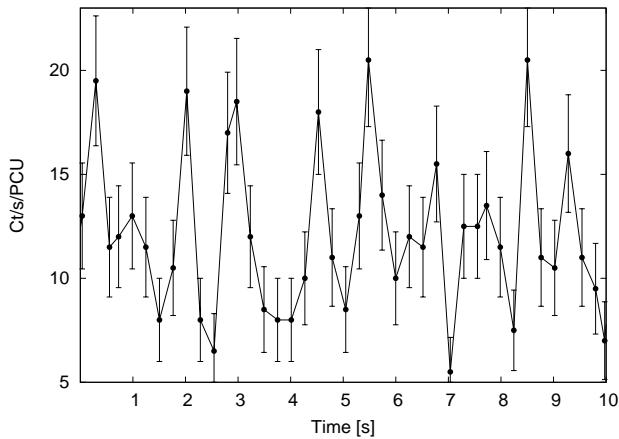}}
  \end{center}
  \caption{Ten-second $\approx2-16$ keV lightcurve with the 1 Hz modulation
    clearly observed. The lightcurve is created by using Event data
    (ObsId: 94315-01-14-00) to have a sufficient time resolution to
    display the 1 Hz modulation. \label{1Hz}}
\end{figure}

We study the energy dependence of the 1 Hz modulation by selecting the
observation with the highest signal-to-noise (ObsId 94315-01-14-00)
and splitting the data into four energy bands (2.47$-$5.31, 5.71$-$9.81,
10.22$-$15.60 and 16.01$-$28.67 keV). The intrinsic faintness of NGC
6440 X-2 does not allow a splitting of the data into more sub-bands. We
do not select any energy above $\approx28$ keV as the count rate is
consistent with the background.

We also investigate the coherence of the time variability in different
energy bands, closely following the method outlined in \citet{vau97}
and \citet{sca13}. The coherence between two time series is defined
as the degree of linear correlation between two Fourier frequencies in
two different lightcurves observed simultaneously but in different
energy bands.  Once each lightcurve is corrected for the Poissonian
noise component, then it is possible to calculate the
\textit{intrinsic} coherence function $\gamma^2_{int}$.  In simple
terms, the intrinsic coherence function describes how strongly the
variability observed in one energy band is causally connected to the
variability seen in the other band.  If the the degree of coherence is
high, then the variability seen in one energy band is
correlated to the variability observed in the other band.

In our measurements, we use again the observation 94315-01-14-00
($\approx9$ ks exposure) and define two energy bands: 2.47$-$7.76 keV
(soft band ``x'') and 8.17$-$28.67 keV (hard band ``y''). The two
bands are defined so that the counting statistics is similar in both
time series. We split the two time series in $N=33$ data segments of
length 256 s each.  Each $i$-th time series $x_i(t)$ and $y_i(t)$
(with $i = 1,...,N$) is Fourier transformed into $X_i(\nu)$ and
$Y_i(\nu)$.  We then calculate the cross-spectra between each pair as:
\begin{equation} 
C_{i}(\nu)=X_i^*(\nu) Y_i(\nu)
\end{equation}
and the power spectral density of each time series: $P_x(\nu) =
|X(\nu)|^2$ and $P_y(\nu) = |Y(\nu)|^2$

 In an ideal linear system with an input time series $x(t)$, the
 output time series $y(t)$ will be perfectly correlated to $x(t)$
 provided that no extra noise enters in the measurements. In this case
 the intrinsic coherence function has a value of 1. Two completely
 unrelated time series have instead an intrinsic coherence value of 0.
 The intrinsic coherence function is obtained by removing the
 Poissonian noise level $n$ from each power spectrum and correlation
 function.  We then define the intrinsic coherence by following
 \citet{sca13}:
\begin{equation}
\gamma^2_{int}(\nu) = { {|\langle \langle C(\nu) \rangle \rangle | ^2 -n^2}\over {\langle \langle |X(\nu)|^2 \rangle \rangle} {\langle \langle |Y(\nu)|^2 \rangle \rangle} },
\label{eq:1}
\end{equation}
where the double brackets refer to averages over the N ensembles and
the number of adjacent Fourier frequency bins.

\section{Results}
\label{sec:results}

We detect the 1 Hz modulation in seven distinct Obs-Ids and six
different outbursts (30-Aug-2009, 21-Mar-2010, 12-Jun-2010,
04-Oct-2010, 23-Jan-2011, 06-Nov-2011; see Figure~\ref{lc} and Table
1), with power in a narrow range of spectral frequencies ($\sim0.01-10
$Hz). In the remaining four outbursts we cannot place any significant
constraint on the presence of the 1 Hz modulation. Indeed one outburst
was observed with the \textit{Swift} X-Ray Telescope (XRT) in photon
counting mode (2.5 s time resolution) with a Nyquist
frequency of only 0.2 Hz. The remaining three outbursts (30-Jul-2009,
01-Oct-2009, 28-Oct-2009) were observed within the PCA Galactic Bulge
Scan program~\citep{swa01} and analyzed in~\citet{alt10}. In these
latter three outbursts the power spectra show no significant feature
due to the poor counting statistics dominated by photon counting
noise.

\begin{figure}[!th]
  \begin{center}
   \rotatebox{0}{\includegraphics[width=1.0\columnwidth]{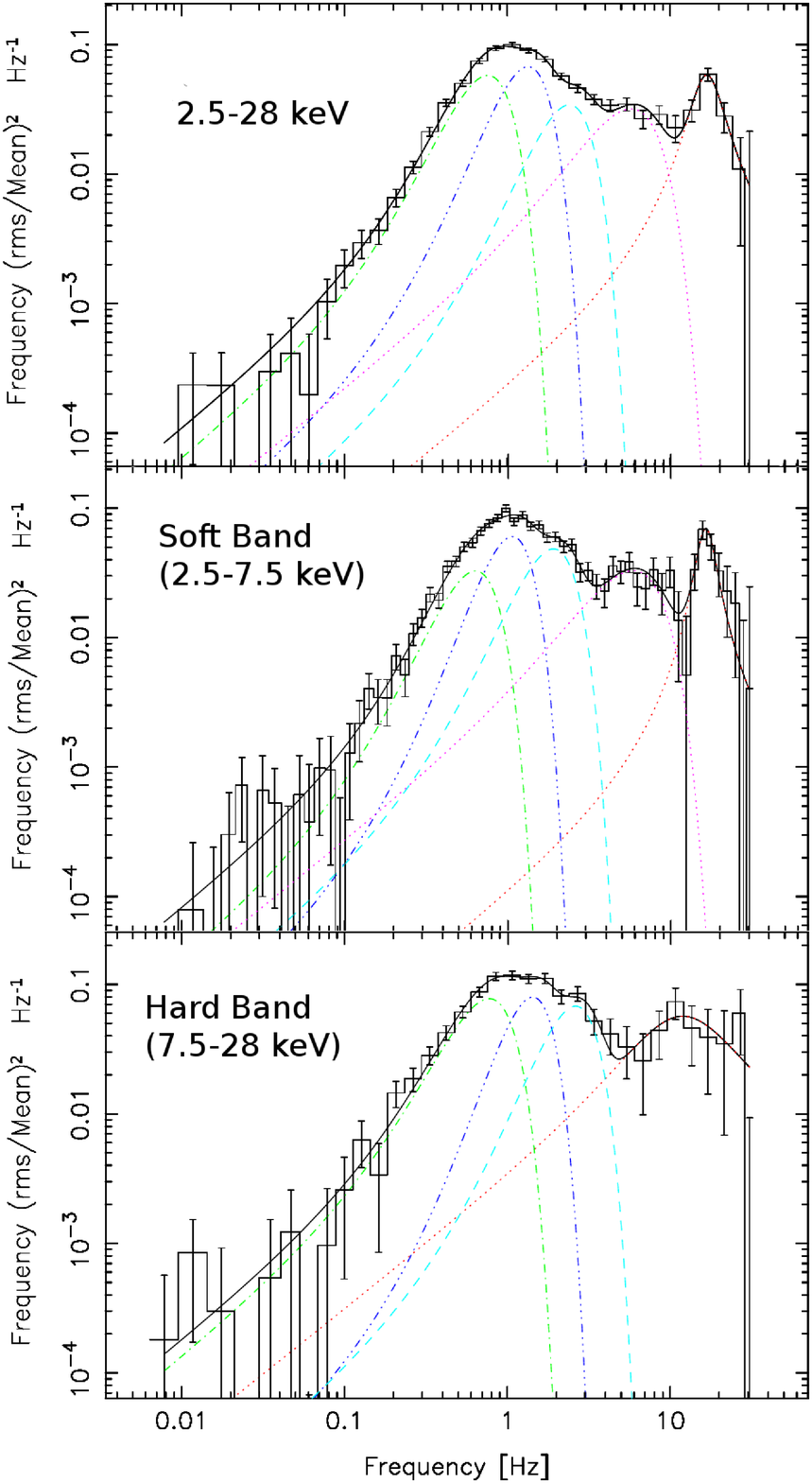}}
  \end{center}
  \caption{Power spectrum of the observation 94315-01-14-00 in the
    broad energy band (top panel), soft band (middle panel) and hard
    band (bottom panel). The 1 Hz modulation appears in the
    $\approx0.1-10$ Hz band with a peak at about 1 Hz and can be
    modeled with a Gaussian plus several harmonics. A higher frequency
    QPO appears at $\approx16$ Hz and is seen with more evidence in
    the soft band, whereas it turns into a broad feature at higher
    energies.\label{power}}
\end{figure}

In some observations the 1 Hz modulation is also clearly observed in
the time domain (see Figure~\ref{1Hz} and compare to Figure 4 in
\citealt{pat09c}). The 1 Hz modulation appears as a Gaussian in the
power spectrum along with one or more other Gaussians whose centroid
frequency is harmonically related to the ``fundamental'' Gaussian
centroid frequency (see Figure~\ref{power}). The quality factor of the
Gaussians is rather low, with $Q=\nu_0/FWHM\simeq 1-2$ and the
amplitude is 40-55\% rms (2-60 keV full energy band) in three
observations. In the remaining observations the feature turns into a
broadband-limited noise feature with rms amplitude between 20 and 40\%
rms.  In all these observations the X-ray flux is always above 3
mCrab, whereas in all the others with a lower flux, 
the source NGC 6440 X-2 is not detected with a measured flux
consistent with background emission.

In addition to the 1 Hz modulation, in two observations we also detect
a broad QPO ($Q\simeq 1$) with centroid frequency of $7.8\pm1.0$ Hz
(Obs-Id: 94044-04-02-00; reported also in \citealt{alt10}) and
$16.8\pm0.9$ Hz (Obs-Id: 94315-01-14-00). This QPO is well fitted in
the power spectrum with a Lorentzian function.

\subsubsection{Energy Dependence, Coherence and Time Lags}
\label{sec:energy}

\begin{figure}[!th]
  \begin{center}
    \rotatebox{-90}{\includegraphics[width=0.7\columnwidth]{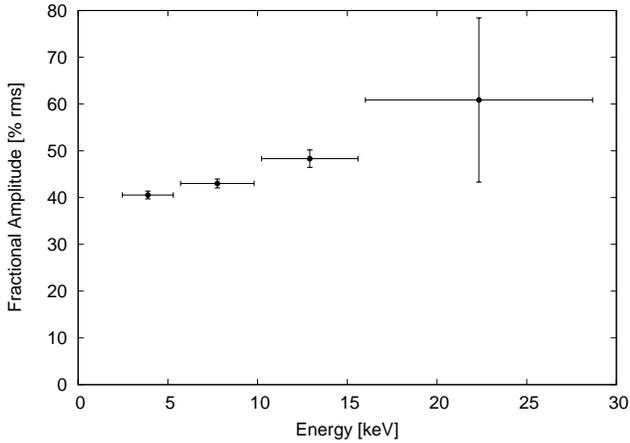}}
  \end{center}
  \caption{Energy dependence of the 1 Hz modulation fractional rms amplitude
(Obs-Id: 94315-01-14-00).
    The fractional amplitude increases with energy from 40\% rms up to
    60\% in the higher energy band. The errorbars on last point
    (16-28 keV) are very large but are consistent with a linear
    increase in the rms amplitude of the 1 Hz modulation. Energies
    above $\approx30$ keV are not included as they are dominated by
    background noise.  The energy dependence closely resembles what
    observed in \SAX~(see \citealt{pat09c}). \label{energy}}
\end{figure}

\begin{figure}[!th]
  \begin{center}
   \rotatebox{0}{\includegraphics[width=1.0\columnwidth]{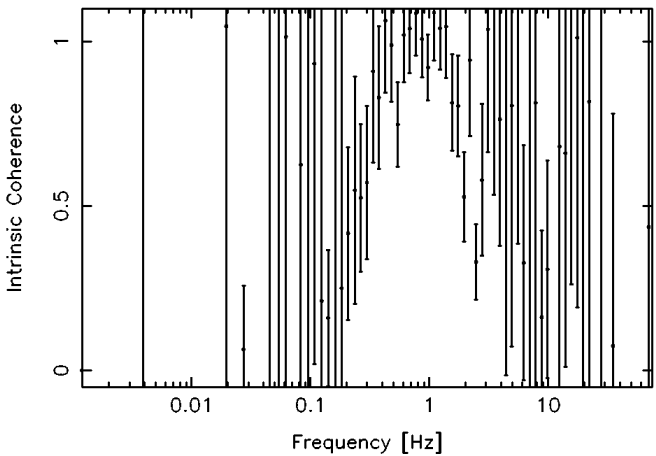}}
  \end{center}
  \caption{Intrinsic coherence function between energy bands
    2.47$-$7.35 keV and 7.76$-$28.67 keV for the observation
    94315-01-14-00 (the observation with the highest signal-to-noise
    ratio). The 1 Hz modulation appears in the $\approx0.1-10$ Hz band
    with a peak at about 1 Hz. The intrinsic coherence function shows how
    the 1 Hz modulation appearing in the two X-ray lightcurves in the
    two energy bands selected is causally connected (intrinsic
    coherence $\approx 1$).\label{intcoherence}}
\end{figure}

We detect a clear increase of the 1 Hz modulation rms amplitude at
high energies, with a rms amplitude that grows linearly from 40\% up
to a peak of $\simeq60\%$ rms in the 16-28 keV energy band
(Figure~\ref{energy}). The intrinsic coherence between two broad
energy bands (2.47$-$7.35 keV and 7.76$-$28.67 keV) is $\approx1$ at
frequencies between 0.5 and 1.5 Hz, and drops substantially outside this 
frequency range. Above 3 Hz and below 0.1 Hz the statistics become poor and the
data points in the intrinsic coherence function have large error bars
(Figure~\ref{intcoherence}).  We use the cross-spectra to inspect for
the presence of time lags between the soft and hard energy bands.  The
time lags are defined as:
\begin{equation}
\tau(\nu) = \frac{\Phi(\nu)}{2\pi\nu}= \frac{arg\left[\langle \langle C(\nu) \rangle \rangle\right]}{2\pi\nu}
\end{equation}
where $\Phi(\nu)$ is the Fourier phase lag between two identical Fourier
frequencies ($\nu$) measured between the soft and hard energy bands. 
No time lags are detected between the two time series, with typical upper
limits on the time lags between a specific Fourier frequency of less
than 5 ms.

\subsection{Pulsations}

We detect significant pulsations at a frequency of 
$205.892$ Hz in all seven Obs-Ids. The frequency shows a Doppler modulation
consistent with the orbital solution reported in \citet{alt10}, although
we do not attempt to refine the solution with a coherent timing analysis.
The pulsations are detected in coincidence with the presence of the
1 Hz modulation. 

\section{Discussion}
\label{sec:discussion}

The spectral and timing properties of NGC 6440 X-2 provide a wealth of
observations to compare with the predictions of the trapped disk and
disk instability model presented in \citet{dan10,dan11,dan12}. Below we
argue that the 1 Hz modulation spectral energy dependence and
coherence between energy bands suggest the variability is driven by a
changing accretion rate onto the star, while the 1 Hz modulation timescale,
amplitude and mean accretion rate are in line with the predictions
from \citet{dan10,dan12}. We also discuss the detailed timing
properties of the source: the unusual power spectral shape of the modulation,
 and the short recurrent outbursts. 

\subsection{Evidence for accretion rate modulation}
\label{sec:mdot_change}

Our analysis of the outbursts of NGC 6440 X-2 shows that the 1 Hz
modulation is present with high amplitude in at least six out of ten
outbursts and that it is a recurrent phenomenon. The lack of the 1 Hz
modulation in the remaining four outbursts is unconstrained due to the
poor photon flux or time resolution limits. The timing and energy
properties of the outburst strongly suggest that the modulation is
produced by variations in the accretion rate onto the star.

Coherent pulsations at 206 Hz are clearly detected during all
outbursts, demonstrating that the channeled accretion flow persists
when the 1~Hz modulation is seen. A complete study of the periodic variability
of the pulsations will be presented elsewhere. 

The energy spectrum~\citep{hei10} and energy-timing information (this
work) also suggests a large-amplitude variation in accretion rate onto
the star at one second intervals. The spectral energy distribution
across all outbursts can be fit with a power law with an index between
$\Gamma\approx 1.7-3.4$~\citep{hei10}, which is most likely produced by inverse
Compton scattering of thermal seed photons, both from hot
accretion-shocked gas on the stellar surface and hot optically thin
gas in the inner accretion flow.

The soft and high energy bands are strongly coherent between 0.5--1.5
Hz, consistent with an overall change in luminosity produced by a
changing accretion rate onto the star, where the variability timescale
is long enough for the optically thin corona above the neutron star and
thermal emission from the neutron star surface to reach
quasi-equilibrium \citep{haa91}.  The drop in coherence
above this frequency may then be a reflection of the intrinsic timescale
for density or temperature fluctuations in the hot corona.

We also measure a steep increase (by about 20\%) of the 1 Hz
modulation rms amplitude with energy, so that the flux at higher
energies varies more than the flux at low energies. Assuming that the
overall emission is dominated by the accretion shock right above the
neutron star's surface (see e.g., \citealt{gie05}), the increased rms
amplitude at higher energies naturally results from a changing accretion
rate. 

Changes in the density and temperature in an optically thin corona
above the neutron star will lead to a variable slope of the power-law
index in the spectrum. (The slope is set by the Compton $y$ parameter,
a function of the gas's temperature and optical depth, where higher
temperatures and higher densities both produce harder spectra). A
fluctuation in accretion rate will thus change not only the overall
flux but also the hardness of the inverse Compton spectrum, so that
flux at higher energies is more variable than at lower energies, as
seen in both NGC 6440 X-2 and \SAX.

No thermal component was detected in the 49.1ks $Chandra$ observation
during the July 28, 2009 outburst \citep{hei10} or any other (less
constraining) observation. This is in contrast to \SAX, in which a
thermal component is frequently seen in outbursts, and is attributed
to a hot spot on the surface of the neutron star
\citep{pou03,gie05,pat09e,pap09,kaj11}. In \SAX\, the thermal
component is observed to lag the high energy one by $\sim0.125$ ms
\citep{cui98,gie02}, which \citet{gie05} attribute to strong Doppler
beaming of the blackbody emission compared to the more diffuse inverse
Compton emission. The lack of thermal emission and low pulsation
fraction may indicate that the main accretion hot spot in NGC 6440 X-2
is beamed out of our line of sight, while the upper limits on phase
lags between hard and soft energy bands (of the order of 5 ms) are not
particularly constraining.

\subsection{Oscillations in an unstable dead disk}
\label{sec:oscillations}

The instability studied in \citet{spr93} and \citet{dan10,dan12} appears
when the inner edge of the disk moves into a marginally dead state,
with $r_{\rm m} \simeq r_{\rm co}$. This condition requires a low mean accretion
rate so that the inner edge of the disk moves temporarily outside the
corotation radius. NGC 6440 X-2, which has one of the lowest outburst
luminosities of all known AMXPs, supports this picture. Its typical
X-ray outburst luminosity is $\approx10^{36}\ergs$\citep{hei10},
corresponding to a mass accretion rate of $\approx
2\times10^{-10}\msun\rm\,yr^{-1}$ for a canonical neutron star mass of
1.4$\msun$ and radius of 10 km. 

The disk truncation radius \rin\ is set by the magnetic field, and is
subject to considerable uncertainty due to the uncertainty in exactly
how the disk and field interact. We follow \citet{spr93} and
\citet{dan10} and define the inner disk radius as where the field is
strong enough to force the gas to co-rotate with the star:
\begin{eqnarray}
r_{\rm m} &\equiv& 47.7 \rm km
\left(\frac{0.1}{B_\phi/B_*}\right)^{1/5}\left(\frac{B_*}{4\times10^8\rm
    G}\right)^{2/5}\left(\frac{M_*}{1.4 M_\odot}\right)^{-1/10}\\
\nonumber &&\left(\frac{R_*}{10\rm km}\right)^{6/5}\left(\frac{\dot{M}}{2\times10^{-10}M_{\odot}
  \rm yr^{-1}}\right)^{-1/5}\left(\frac{P_*}{4.9\rm ms}\right)^{3/10},
\end{eqnarray}
where the subscript ``*'' identifies quantities referred to the
neutron star and $B_\phi/B_*$ is the fractional size of the toroidal
field induced by the relative rotation between the magnetic field and
the disk. For $B_\phi/B_* > 1$, the field lines will rapidly inflate
and sever the connection between the disk and the star
(e.g., \citealt{aly85,lov95}), but the exact average value is
uncertain. Assuming $B_\phi/B_* \sim 0.1-1$ implies an uncertainty of
up to $\sim 60\%$ on the value of $B_*$ that will put $r_{\rm m}
\simeq r_c$.  The inferred magnetic field for NGC 6440 X-2 (assuming
$B_\phi/B_* = 0.1$ and $r_{\rm m} = r_{\rm co}$ for $\dot{M}
\sim10^{-10}\msun\,yr^{-1}$) will be $4\times10^8\rm G$, well in line
with most AMXPs (see e.g., \citealt{pat12d} for a review).

The timescale of the 1~Hz modulation also suggests an instability in
the inner accretion disk. This timescale is much longer than the spin
frequency of the star (206~Hz) or the dynamical timescale of the inner
disk (similar when $r_{\rm m} \sim r_{\rm co}$). It is also likely much
longer than the timescale on which the field lines are twisted open
and reconnect to the disk, although this is currently somewhat
uncertain (e.g., \citealt{rom09}). 

The disk instability studied by \cite{spr93} and \cite{dan10,dan12}
causes density fluctuations in the inner parts of the disk, which
generally evolve on viscous timescales. The instability
frequency roughly scales with the viscous timescale around the inner
edge of the disk. This timescale is given by:
\begin{eqnarray}
T_{\rm visc} & \sim & 80 \left(\frac{0.1}{\alpha}\right)^{-4/5}
\left[\frac{\dot{M}}{2\times10^{-10}~{M_\odot \rm yr^{-1}}}\right]^{-3/10}\nonumber\\ &&
\times \left[\frac{M_*}{1.4~M_\odot}\right]^{1/4} \left[\frac{r}{50~ {\rm
      km}}\right]^{5/4}\rm\,s
\label{tvisc}
\end{eqnarray}
where $\alpha\sim0.1$ is the canonical $\alpha$-disk parameter for
geometrically thin and optically thick disks~\citep{siu77} and $r\sim
r_{\rm m}$. For NGC 6440 X-2, $T_{\rm visc} \sim 80\rm\, s$, and the
modulation at 1 Hz is well in the instability range seen by
\cite{dan10,dan12}. In section \ref{sec:disk-field}, we discuss this
further and place tighter physical constraints on the
disk-magnetosphere interaction based on the instability properties.

\subsection{A dead disk in quiescence?}

As well as being weak with short durations, the outbursts seen in NGC
6440 X-2 have the shortest recurrence time of any known AMXP, with ten
outbursts seen at roughly monthly intervals between 2009 and
2011. Similar behaviour was seen in another AMXP, IGR J00291+5934,
which in 2008 underwent two small outbursts (lasting around 10 days
each) separated by a month in quiescence \citep{pat10b, har11}. This
AMXP also showed a QPO at 0.5~Hz~\citep{har11}, although this QPO is
much more typical (with substantially lower amplitude of $(13\pm2)\%$,
and a standard Lorentzian shape) than those seen in NGC 6440 X-2 and
\SAX. There also may be a connection between the appearance of the
instability in \SAX\, and the 'reflaring' portion of the
light-curve. The 'reflaring' refers to large luminosity variations (up
to 3 orders of magnitude) on timescales of 3-5 days in the decay tail
of \SAX\, (see e.g., Figure 1 in \citealt{pat09c}), suggesting
large-scale changes in the accretion rate as the source approaches
quiescence.

The short variation timescale of the reflares and outbursts is
incompatible with the standard ionization instability model believed
to power normal outbursts (e.g., \citealt{las01}) and suggests that a
substantial amount of gas remains in the inner regions of the
accretion disk (close to \rc) when the accretion rate decreases,
exactly as is expected for a dead disk. As was first noted by
\cite{dan12}, the unusual frequent outbursts and strong 1 Hz
modulation are both well-explained by a disk that never moves into a
strong propeller phase at very low $\dot{M}$, but instead stores mass as a
`dead disk', which becomes unstable at higher $\dot{M}$.

The density structure of a dead disk, with its low temperature and a
substantial amount of stored mass in the disk inner regions, is
considerably different from the standard accretion disk and could
strongly alter the ionization instability cycle. Although detailed
calculations of how the ionization instability would be triggered in a
dead disk are beyond the scope of the current work, the additional
stored mass during quiescence could lower the density threshold for
which newly accumulated mass (from the star or the outermost regions
of the accretion disk) can trigger a new outburst. Such outbursts
would be expected to have both shorter recurrence times and weaker
outbursts, since less additional mass is accumulated between outbursts.

A further prediction of the dead disk picture is that spin-down should
be enhanced during the quiescent phase, since the disk inner edge
remains close to \rc\, and continues to spin down the star.  Note
however, that the disk-field coupling parameters we infer for this
source (in section \ref{sec:disk-field}) imply a narrow connected
region between the star and the disk, which will also limit the
spin-down from this source.

\subsection{Constraining properties of the disk-field interaction}
\label{sec:disk-field}

In the model presented in \citet{dan10,dan12}, the
instability develops in two different regimes, depending mainly on the
mean mass accretion rate through the disk. The more-studied regime
(called RI in \citealt{dan12}) occurs at very low accretion rates,
with $\dot{M} \approx 10^{-4}-0.1 \dot{M}_{\rm co}$ (where
$\dot{M}_{\rm co}$ is the accretion rate that sets $r_{\rm co} = r_{\rm
  m}$). It is characterized by long periods of quiescence while mass
accumulates in the inner regions of the disk followed by brief
accretion outbursts. Indeed, the authors found that the period of the
instability could span several orders of magnitude, from around
$0.1-10^4\, T_{\rm visc}$ (where $T_{\rm visc}$ is the viscous timescale
at $r_{\rm m}$, Equation \ref{tvisc}), depending chiefly on the time needed to fill up
the reservoir and tip the source into an accretion burst.

\begin{figure}[!t]
  \begin{center}
    \rotatebox{0}{\includegraphics[width=1.0\columnwidth]{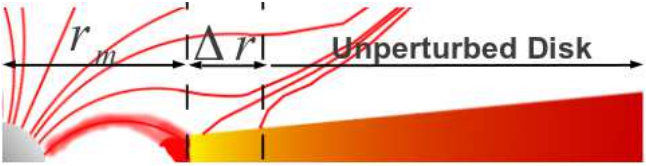}}
  \end{center}
  \caption{Illustration of the interaction region $\Delta\,r$ where
    the neutron star and disk magnetic fields are coupled. At $r<r_m$
    the neutron star magnetic field lines are closed, whereas at
    $r_m<r<r_m+\Delta\,r$ the field lines open and interact with the
    accretion disk.  Beyond the interaction region the disk flow is
    unperturbed.\label{disk}}
\end{figure}

\begin{figure}[!t]
  \begin{center}
    \rotatebox{0}{\includegraphics[width=1.0\columnwidth]{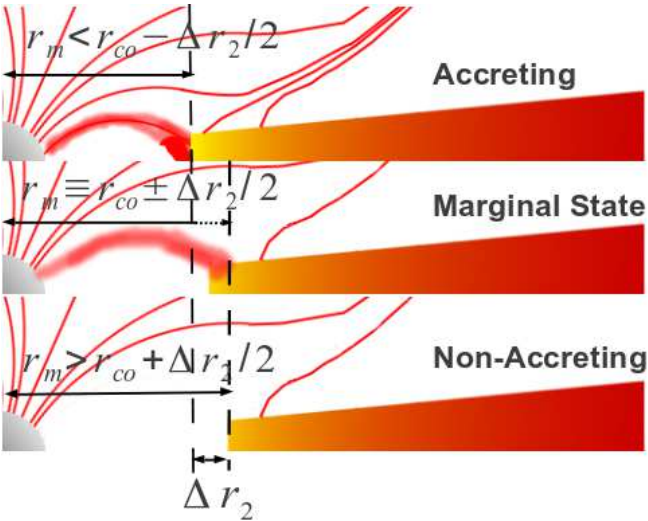}}
  \end{center}
  \caption{Transition region $\Delta\,r_2$ around \rc. When \rin\, is
    smaller than $r_{\rm co}-\frac{\drr}{2}$ the gas accretes along
    magnetic field lines, and spins up the star. When \rin\, lies in
    the transition region $r_{\rm co}\pm\frac{\Delta\,r_2}{2}$ the
    disk is in a marginal state, where accretion onto the star
    continues but the inner disk simultaneously begins extracting
    angular momentum from the star, so that the spin up turns to spin
    down. When \rin\, lies above $r_{\rm co}+\frac{\Delta\,r_2}{2}$
    the accretion flow stops but the inner disk radius remains close
    to the corotation radius. When the 1 Hz modulation is seen, the
    inner disk edge stays around $r_{\rm m}\simeq r_{\rm
      co}\pm\frac{\Delta\,r_2}{2}$. \label{disk2}}
\end{figure}

The instability outburst shape and duration also depends quite
sensitively on the details of the disk-field interaction, which
\cite{dan10,dan12} parametrized as two length-scales: the width of
the coupled region between the field and the disk, $\dr$, and $\drr$,
the transition length around \rc\, between the accreting and
non-accreting solutions (see Figure~\ref{disk} and ~\ref{disk2}). 
Both of these length scales are assumed to be
small: $\dr/ r_{\rm co}, ~\drr /r_{\rm co} \ll 1$. In the RI
instability region, \cite{dan10,dan12} found that the instability
occurs for a broad range of $\dr/r_{\rm m}$ (see Figure 3 and 4 in
\citealt{dan12}) but only when $\drr/r_{\rm m}$ is very small (of
order of 0.01).

However, in \cite{dan12} the authors identified a second instability
region, which they referred to as RII. Here the mean accretion rate is
$0.1-10\times\dot{M}_{\rm co}$ -- much higher than RI -- and accretion
continues throughout the cycle, so that the amplitude of the
instability is much smaller. The instability timescale is also much
shorter than for RI, with periods around $0.01-0.1\,T_{\rm visc}$, and
has a more sinusoidal, regular shape and a smaller amplitude,
although the outburst length is still typically shorter than the build-up
time (see e.g., Fig. 5 of \citealt{dan12}). The instability
also appears for different disk-field interaction parameters: here
$\dr/r_{\rm m}$ is very small ($\approx 0.05$), while $\drr/r_{\rm
  m}$ has a much broader range of instability.
  
In NGC 6440 X-2 the corotation radius is located at approximately 50
km from the neutron star center, but the surface magnetic field is
still unknown, so that it is not possible to estimate the location of
the magnetospheric radius from the luminosity alone. However, if we
assume a typical AMXP magnetic field strength of $\approx(1-5)\times
10^8$ G, the observed mass accretion rate is of the same order of
magnitude or lower than $\dot{M}_{\rm co}$, suggesting that we are
observing the RII instability. In \SAX\, (with $B\approx 10^8$ G, see
\citealt{har08,pat12}), the 1 Hz modulation was seen at fluxes
corresponding to $\dot{M}/\dot{M}_{\rm co}\approx 1$, suggesting that
source is also unstable in the RII region. In this regime the
instability only appears over a relatively narrow range of $\dot{M}$,
offering a plausible explanation for why the instability is only
observed in the decay tail of the outburst.

One obvious point of interest is why the 1 Hz modulation central
frequency is nearly the same in both \SAX\, and NGC 6440 X-2, given
that the system parameters are quite different. Despite the comparable
X-ray flux detected in both AMXPs when the 1 Hz modulation is
observed, the corresponding mass accretion rate of \SAX\, is about an
order of magnitude lower than in NGC 6440 X-2 because the first source
lies at about 3.5 kpc~\citep{gal06} whereas the latter is at 8.5
kpc~\citep{ort94}.  If we compare the fundamental centroid frequency
of the 1 Hz modulation observed in all observations of NGC 6440 X-2
with those of \SAX\, then it is, on average, systematically lower in
NGC 6440 X-2 by only a factor of $\approx 2$.  Without knowing the
magnetic field in either source it is not easy to constrain exactly
where the disk is truncated (or, to invert the problem, the actual
ratio $\dot{M}/\dot{M}_{\rm co}$), but if we assume that the relevant
timescale is the viscous timescale at \rc\, then this will be $\sim
80$ s for NGC 6440 X-2 vs. $\sim 40$ s for \SAX, so the model predicts
roughly the same $\nu$ for both sources, as observed. An
independent constraint on $B_*$ would allow a much more secure
determination of $\dot{M}/\dot{M}_{\rm co}$ and also $\drr$ and $\dr$.

The RII instability appears only for very small values of $\dr$, of
order 0.01--0.05 $r_{\rm m}$. This suggests a very small coupling
region between the field and the disk -- of order 1 km -- in both \SAX\,
and NGC 6440 X-2. Given the uncertainties in determining the system's
physical parameters, $\drr$ cannot be as well-constrained, although at
these accretion rates $\drr \propto \nu^{-1}$ (where $\nu$ is the
instability frequency; see Fig. 7 of \citealt{dan12}), so a smaller value
for $\drr$ (of about 1-10 km) would also seem to be favored.

\citet{dan12} found that when the instability
is triggered in the RII regime, the accretion torques are reduced
compared with the standard disk solution.  An implication of this is
that if an AMXP is observed to show the 1 Hz modulation in a certain
mass accretion rate range, then the observed accretion torques must be
smaller than the prediction made with the approximation:
\begin{equation}
N = \mdot\sqrt{G\,M_{*}\,r_{\rm m}}
\end{equation}
It is possible to speculate therefore that the lack of strong positive
accretion torque in \SAX\, \citep{har08, har09} might be partially
caused by the source being almost always close to this marginally
accreting state. The fact that \SAX\, is found with $r_{\rm m}$ close
to $r_{\rm co}$ was also suggested by \citet{has11} based on coherent
timing and quiescent thermal properties of the accreting neutron star.

\subsection{Detailed timing properties}
\label{sec:timing_prop}

The fact that the 1 Hz modulation can be fitted with a Gaussian and not
with a Lorentzian suggests some intrinsic difference between the 
generation of the 1 Hz modulation and the higher frequency QPOs observed
in NGC 6440 X-2 and \SAX. It also suggests some remarkable difference
between most of the other QPOs broadly observed in other low mass X-ray binaries
and the 1 Hz mechanism. 

If a Gaussian stochastic process controls the fluctuations in mass
accretion rate around the magnetospheric radius, then the expected
$\mdot$ profile instability will be a Gaussian. This in turn might
explain why the 1 Hz modulation has a Gaussian shape and not a Lorentzian. It is
conceivable indeed to think that, once the instability is triggered
and the gas is accumulating around $r_{\rm m}\approx r_{\rm co}$, the
main timescale of the instability will be strictly periodic if none of
the quantities in the system is changing (i.e., $\mdot$, $\dr/r_{\rm
  m}$ and $\drr/r_{\rm m}$).  However, if we let $\mdot$ fluctuate (in
a Gaussian stochastic way) around its mean value, the periodicity of
the instability will slightly change, thus broadening the periodicity
across adjacent frequencies. It is then possible to understand 
why the shape of the 1 Hz modulation is Gaussian and why it
becomes a broad feature in some observations whereas in others it is a
more periodic and narrow feature: the stronger the fluctuations in
$\mdot$ around the mean value, the broader the 1 Hz modulation will
appear in the power spectra.

\subsection{Alternative Models}

\citet{pat09c} reported an extensive discussion of alternative models
to the trapped-disk instability presented in 
\citet{spr93} (and later developed in \citealt{dan10,dan12}). It was shown that none of the alternative models can
explain at the same time the observed features of the 1 Hz modulation
in SAX J1808.4--3658. Identical consideration apply also to the 1 Hz
modulation in NGC 6440 X-2, as all the observational features closely
resemble those in SAX J1808.4--3658 and identical considerations
apply. More recently, \citet{rom13} has carried 3D MHD simulations
with a dipole magnetic field misaligned from the neutron star's
rotational axis, in which a warped disk appears. Perturbations of the
disk in the vertical direction produce low-frequency bending waves,
whose frequency is approximately equal to the Keplerian frequency
$\left(\Omega=\sqrt{GM/r^3}\right)$ of the gas in the disk.  Since the
inner disk is truncated at about $\sim50$ km, the Keplerian frequency
there is of the order of 200 Hz, which is 2-3 orders of magnitude off
the observed instability frequency.  The warped disk predicts
oscillations with frequency down to 0.1-1 Hz, but the region of the
disk that oscillates has to be far away from the neutron star, at
several thousands kilometers, where no X-ray radiation is produced
and the inner accretion flow (at about 50 km) would still be modulated
at the much higher Keplerian frequency there.

\section{Conclusions}
\label{sec:conclusions}
We have studied the 1 Hz modulation in the AMXP NGC 6440 X-2 and
suggested that this feature has the same physical origin as seen in
the other AMXP \SAX\,.  We propose the identification of this feature
with a magnetospheric instability triggered in the weak propeller
stage.  Our study suggests a very small transition region $\dr/r_{\rm
  m}$ of the order of 0.01--0.05.  This region is where the magnetic
field lines open and interact with the accretion disk and this has to
be small in both AMXPs considered in this work.  We are not able to
put such strong constraints on transition region $\drr/r_{\rm m}$,
where the nature of the accretion torques changes from spin up to spin
down, but suggest it lies in the range 0.01-1. Given that $r_{\rm
  m}\simeq r_{\rm co}$ when the instability is observed, the radial
extension of the two transition regions for the two AMXPs is:
$\dr\approx 1$ km and $\drr\sim1-10$ km. This is the first time that
the width of the magnetosphere-disk interaction region can be
constrained with direct observations.

We have also suggested an explanation for the Gaussian shape of the 1
Hz modulation as seen in the power spectra as due to fluctuations in
the mean mass accretion rate during the instability. Strong $\mdot$
fluctuations can smear the 1 Hz modulation from a rather coherent QPO
to a broadband-limited noise.  If confirmed, this might be the first
clear physical identification of the mechanism behind the generation
of a low frequency QPO and broad band noise.

\acknowledgments{AP acknowledges partial support from an NWO Veni
  fellowship. CD'A is supported by an NWO Vidi Fellowship (PI
  Watts). We would like to thank Phil Uttley, Brynmor Haskell and Ed
  Brown for inspiring discussions and M. van der Klis, A. L. Watts and
  P. Bult for interesting comments.}


\end{document}